\documentclass[final,3p,sort&compress]{elsarticle}

\usepackage{graphicx}

\usepackage{amsmath,amssymb,amsfonts}
\usepackage{bm}

\def\be{\begin{equation}}
\def\ee{\end{equation}}
\def\ba{\begin{eqnarray}}
\def\ea{\end{eqnarray}}
\def\bd{\begin{displaymath}}
\def\ed{\end{displaymath}}
\def\bq{\begin{eqnarray}}
\def\eq{\end{eqnarray}}

\journal{Annals of Physics}

\begin{document}

\begin{frontmatter}

\title{On the unique mapping relationship between\\ initial and final
quantum states}

\author{A. S. Sanz$^{1,2}$\corref{corresp}} \ead{asanz@iff.csic.es}

\author{S. Miret-Art\'es$^1$}

\cortext[corresp]{Corresponding author}

\address{$^1$Instituto de F\'{\i}sica Fundamental (IFF--CSIC),
Serrano 123, 28006 Madrid, Spain}

\address{$^2$Department of Physics and Astronomy, University College
London, Gower Street, London WC1E 6BT, United Kingdom}

\begin{abstract}
In its standard formulation, quantum mechanics presents a very
serious inconvenience: given a quantum system, there is no
possibility at all to unambiguously (causally) connect a particular
feature of its final state with some specific section of its initial
state. This constitutes a practical limitation, for example, in
numerical analyses of quantum systems, which often make necessary
the use of some extra assistance from classical methodologies. Here
it is shown how the Bohmian formulation of quantum mechanics removes
the ambiguity of quantum mechanics, providing a consistent and clear
answer to such a question without abandoning the quantum framework.
More specifically, this formulation allows to define probability
tubes, along which the enclosed probability keeps constant in time
all the way through as the system evolves in configuration space.
These tubes have the interesting property that once their boundary
is defined at a given time, they are uniquely defined at any time.
As a consequence, it is possible to determine final restricted (or
partial) probabilities directly from localized sets of (Bohmian)
initial conditions on the system initial state. Here, these facts
are illustrated by means of two simple yet physically insightful
numerical examples: tunneling transmission and grating diffraction.
\end{abstract}



\end{frontmatter}



\section{Introduction}
\label{sec1}

Modern-day experimental techniques allow us to monitor the evolution
of quantum systems in real time ---e.g., the passage of electrons
through barriers \cite{uiberacker:Nature:2007}, the formation of
diffraction patterns with complex molecules
\cite{arndt:NatCommun:2011}, or determining the (average) path
followed by photons in Young's two-slit experiment without
destroying the interference pattern \cite{kocsis:Science:2011}. This
possibility brings in a natural way the question of whether it would
also be possible to establish an unambiguous connection between some
particular feature of the final state of a quantum system and a
specific subregion of its initial state. In principle, there are no
means in our standard version of quantum mechanics to determine such
an information in a unique way. Schr\"odinger's equation describes
how the whole quantum state of a system, specified by a wave
function $\Psi({\bf r},t)$, evolves in time, but it is unable to
provide an answer on how a particular piece of this wave function
evolves or to see its effect on the final outcome.

In classical mechanics, such a question is not a problem at all. The
connection can be established because, apart from statistical
density distributions, one can also monitor the system along
trajectories. The evolution of any phase space point (representing
the system) can be monitored in time by means of a well-defined
trajectory. These sets of trajectories, starting from a certain
phase-space region $\Omega_0$, evolve according to a Liouvillian
dynamics. There is a phase-space volume-preserving transformation
that uniquely carries the initial conditions contained in $\Omega_0$
onto a region $\Omega_t$ at a time $t$. That is, both regions are
causally connected through a mapping or flow transformation
$\Omega_t = \Phi(\Omega_0)$
\cite{gutzwiller:1990,sanz:SSR:2004,sanz:PhysRep:2007}. In terms of
statistical distributions, this translates into the continuity
equation
\be
 \frac{\partial \rho}{\partial t} + \nabla \cdot {\bf J} = 0 ,
 \label{ec1}
\ee
known in this context as the Liouville equation. In this equation,
$\rho({\bf x},t)$ represents the probability distribution of
trajectories in phase space (here ${\bf x}=({\bf q},{\bf p})$
denotes collectively the set of generalized positions and momenta
defining this phase space) and a current density with the form ${\bf
J}({\bf x},t) = {\bf F}({\bf x})\rho({\bf x},t)$, where ${\bf F}$ is
defined in terms of the equation of motion describing the
trajectories, $\dot{\bf x} = {\bf F}({\bf x})$. This simple equation
establishes the connection between the variation in time of the
(trajectory) density distribution, $\rho$, contained in $\Omega_t$
at a time $t$ and its flux or current density, ${\bf J}$, across the
boundaries of $\Omega_t$ (henceforth, these boundaries will be
denoted by $\partial\Sigma_t$). Although implicit in the
quasi-classical trajectory method, this simple idea became the germ
for a former methodological scheme aimed at determining reaction
probabilities and product distributions by considering phase-space
{\it reactivity bands}
\cite{pollak:JCP:1980,pollak:JCP:1983,tannor-bk}.

Here we show how to proceed in a similar fashion in quantum
mechanics by making use of its Bohmian formulation. As it is shown,
this hydrodynamical \cite{madelung:ZPhys:1926} or trajectory-based
\cite{bohm:PR:1952-1} formulation offers us a unique answer to the
question posed above. Making use of the non-crossing property in
configuration space of Bohmian trajectories
\cite{sanz:JPA:2008,sanz-bk-1}, we can define {\it probability
tubes}, which are the quantum analog of the classical reactivity
bands mentioned above. The interest of these particular tubes arises
from the fact that, as far as we are concerned, Bohmian mechanics is
the only formulation of quantum mechanics that allows to specify
such elements in a unique way. Once the region $\Omega_0$ is
defined, one has a clear and unambiguous prescription to follow its
causal evolution throughout configuration space. A mapping relation
$\Omega_t = \Phi_Q(\Omega_0)$ (`$Q$' for `quantum map') can also be
used, given that Bohmian trajectories do not cross in configuration
space and obey a continuity equation like (\ref{ec1}) (with ${\bf x}
= {\bf r}$). Thus, trajectories distributed along the boundary
$\partial\Sigma_0$ of $\Omega_0$ will subsequently form the boundary
$\partial\Sigma_t$ of $\Omega_t$. These are {\it separatrix
trajectories} or, in short, {\it separatrices}. We would like to
remark that, somehow, these properties have been empirically
(numerically) observed in the analysis of single features in
realistic simulations, for example, to determine the specific
contribution of each part of the initial wave function to different
diffraction features in realistic simulations of atom-surface
interactions and lifetimes of selective adsorption resonances
\cite{sanz:prb:2000,sanz:JPCM:2002,sanz:jcp:2004,sanz:prb:2004}, to
obtain estimates of product formation rates in model molecular
systems \cite{sanz:cpl:2009,sanz:cpl:2010E}, or in the
implementation of Monte-Carlo-Bohmian samplers employed in quantum
initial value representation calculations
\cite{bittner:JCP:2003,makri:JCP:2003,makri:JPCA:2004}. Furthermore,
the deformation of the tubes with time is not completely arbitrary,
as it usually occurs in topology when dealing with deformations,
since the non-crossing rule of trajectories has to be always
preserved.

A direct consequence of dealing with quantum probability tubes it
that, if we define a {\it partial} or {\it restricted probability}
\cite{sanz:jcp:2005,sanz:cpl:2009,sanz:cpl:2010E,sanz:CP:2011} as
the fraction of the total probability that has ended up inside a
region or domain $\mathcal{D}$ of the system configuration space,
\be
 \mathcal{P}_\mathcal{D} (\infty) \equiv
  \lim_{t \to \infty} \mathcal{P}_\mathcal{D} (t)
  = \lim_{t \to \infty} \int_\mathcal{D} \rho({\bf r},t) d{\bf r} ,
 \label{ec3}
\ee
it will remain constant in time whenever $\mathcal{D}$ corresponds
to a probability tube. This means that, in principle, asymptotic
probabilities like (\ref{ec3}) can be specified from the initial
state without any further calculation if we know: (1) the analytical
form for the separatrices defining the initial boundary
$\partial\Sigma_0$, and (2) any {\it bifurcation} or {\it branching
process} undergone by the probability tubes between $t_0$ and
$t\to\infty$. The presence of branchings is actually a very
important issue: any region $\Omega_t$ (including $\Omega_0$) may
consist of more than one separate subregions, which emerge or
disappear along time.

This work has been organized as follows. The proof of the uniqueness
of the probability tubes, as well as their definition and that of
separatrix, are given in Sec.~\ref{sec2}. The properties of these
tubes are illustrated and analyzed in Sec.~\ref{sec3} in terms of
two simple yet physically insightful one-dimensional numerical
examples: tunneling transmission and grating diffraction (for more
complex applications we refer the interested reader to the
references mentioned above). Finally, in Sec.~\ref{sec4} we
summarize and discuss the main conclusions drawn from this work.


\section{Theoretical background}
\label{sec2}

Let $\mathcal{P}_{\Omega_t}(t)$ be a quantity describing the
time-evolution of a certain probability of interest (e.g., a
reaction probability, a transmittance, a cross-section, etc) inside
a region $\Omega_t$ of the corresponding configuration space. This
quantity is given in terms of the partial or restricted probability
\be
 \mathcal{P}_{\Omega_t}(t) \equiv \int_{\Omega_t} \rho({\bf r},t) d{\bf r} .
 \label{prob}
\ee
The variation of $\mathcal{P}_{\Omega_t}(t)$ with time inside
$\Omega_t$ is
\be
 \frac{d\mathcal{P}_{\Omega_t}(t)}{dt} =
  \int_{\Omega_t} \frac{\partial \rho}{\partial t}\ d{\bf r} ,
 \label{ec4}
\ee
although it can also be written as
\be
 \frac{d\mathcal{P}_{\Omega_t}(t)}{dt}
  = - \int_{\Omega_t} \left( \nabla \cdot {\bf J} \right) d{\bf r}
  = - \int_{\partial\Sigma_t} {\bf J} \cdot d{\bf S} .
 \label{ec5}
\ee
In the second equality of (\ref{ec5}), which is a straightforward
application of the divergence or Gauss-Ostrogradsky theorem, $d{\bf
S}$ denotes a vector normal to a surface element
$d(\partial\Sigma_t)$ of $\Omega_t$ and pointing outwards. By
combining Eqs.~(\ref{ec4}) and (\ref{ec5}), we find that the losses
or gains of $\mathcal{P}_{\Omega_t}(t)$ inside $\Omega_t$ are
described, respectively, by the outgoing or ingoing probability flux
${\bf J}$ through $\partial\Sigma_t$. This is a well-known result,
which translates into the continuity equation
\be
 \frac{\partial \rho}{\partial t}
  + \nabla \cdot \left( {\bf v} \rho \right) = 0 .
 \label{ec1b}
\ee
when generalizing to the whole configuration space and making the
flux to be independent of $\Omega_t$. However, we obtain another
interesting result if we still keep the dependence on $\Omega_t$,
for it also tells us that if the time-evolution of this region
follows some particular rule, then one could keep the value of
$\mathcal{P}_{\Omega_t}(t)$ constant all the way through.

The standard version of quantum mechanics does not give us any
prescription to monitor in time portions of a given quantum system.
Therefore, the choice of a time-evolution rule for $\Omega_t$ is
left to arbitrariness. However, in the Bohmian formulation of
quantum mechanics, the evolution of a system can be monitored by
means of well-defined trajectories in configuration space. These
trajectories are solutions to the equation of motion
\be
 \dot{\bf r} = \frac{\nabla S}{m} = \frac{\bf J}{\rho} .
 \label{ec6}
\ee
where $m$ is the mass associated with the system. In this equation,
$S({\bf r},t)$ denotes the phase of the wave function when it is
recast in polar form, i.e.,
\be
 \Psi({\bf r},t) = \rho^{1/2} ({\bf r},t)\ \! e^{iS({\bf r},t)/\hbar} ,
 \label{ansatz}
\ee
with $\rho = |\Psi|^2$ being the probability density, and ${\bf
J}({\bf r},t)$ is the local quantum current density, defined as
\be
 {\bf J} = \frac{\hbar}{2mi}
  \left[ \Psi^* \nabla \Psi - \Psi \nabla \Psi^* \right] .
\ee
As it can be noticed, Eq.~(\ref{ec6}) is identical to the classical
Jacobi law of motion \cite{goldstein-bk}, although $S$ is not the
classical action---in semiclassical treatments, a relationship can
be found between this quantum phase and the classical action
\cite{sanz:SSR:2004}. In principle, Eq.~(\ref{ec6}) is postulated
once the ansatz (\ref{ansatz}) is substituted into Schr\"odinger's
equation and two couple real equations arise, namely the continuity
equation and the quantum Hamilton-Jacobi equation. However, such a
reformulation will not be necessary here, since (\ref{ec6}) can be
introduced in a natural way as a local velocity field from the
relation ${\bf J} = \rho {\bf v}$, directly connected with the
continuity equation (\ref{ec1b}). Physically, this velocity field
describes how the quantum probability density is transported through
the configuration space in the form of the quantum probability
current density. Notice that, according to Eq.~(\ref{ec6}), this
field remains invariant if the wave function is multiplied by a
time-dependent gauge factor $e^{ig(t)/\hbar}$, but not if such a
gauge depends on the spatial coordinate, i.e.,
$e^{iG(x,t)/\hbar}$. In such a case, there will be a contribution
$\nabla G(x,t)/m$ that will change locally (i.e., at each point $x$)
the velocity field.

Now we already have all the elements to provide an answer to our
question without abandoning the theoretical framework of quantum
mechanics (notice that Bohmian mechanics constitutes another way
to recast quantum mechanics, which must not be seen as a
semiclassical-like approximation). Thus, consider that the
boundary $\delta\Sigma_t$ consists of the positions $x(t)$ at $t$
of a set of Bohmian trajectories. From the deterministic equation
of motion (\ref{ec6}) we can trace these trajectories back and
also propagate them ahead in time. In any case, at any other time
$t'$ these trajectories will form another boundary
$\delta\Sigma_{t'}$. Moreover, as happens with their classical
counterparts in phase space, which cannot cross through the same
point at the same time, Bohmian trajectories are also constrained
to move in configuration space without crossing
\cite{sanz:JPA:2008,sanz-bk-1}. Accordingly, the arbitrariness of
quantum mechanics is thus removed, since we now have a clear and
unambiguous prescription to follow the evolution of a certain
region $\Omega_t$ in time.

The trajectories forming the boundary $\delta\Sigma_t$ play the role
of separatrices, since trajectories outside $\Omega_t$ cannot
penetrate this region, nor those inside can leave it. This means
that the probability inside $\Omega_t$ remains always constant in
time. Hence, we can define quantum probability tubes as the
structures formed by the time-evolution of $\Omega_t$. If the
dimensionality of the problem is $N$ ($N$ is the number of degrees
of freedom), these tubes will have a $(N+1)$-dimensionality. These
tubes are uniquely selected once a specific region of the
configuration space has been chosen. For example, in a scattering
problem, if we define a number $M$ of regions $\Omega_t^{(m)}$ ($m =
1, 2, \ldots M$), each one covering one particular diffraction
feature, when we trace them back in time, we will obtain a full map
of regions $\Omega_0^{(m)}$ covering the initial wave function and
providing us information about the sets of initial conditions that
have given rise to such diffraction features (see, for example,
Ref.~\cite{sanz:prb:2000}).

From the above statements we obtain an interesting consequence: any
restricted probability can be determined directly from the initial
state if the end points of the associated separatrix trajectories as
well as any intermediate branching process are known. That is, in
principle one could determine (or, at least, get an estimate of)
final probabilities without carrying out the full calculation, but
directly from the particular region covered by the initial wave
function causally connected with the feature of interest
\cite{sanz:JPA:2011}. The proof is very simple and follows
straightforwardly from the possibility to define probability tubes.
Consider that in the restricted probability (\ref{ec3}), the domain
$\mathcal{D}$ corresponds to the region $\Omega_\infty$, which is
the asymptotic extreme of a probability tube starting in $\Omega_0$
at $t=0$. By integrating back in time (i.e., considering the inverse
mapping transformation $\Omega_0 = \Phi_Q^{-1}(\Omega_\infty)$), we
find
\be
 \mathcal{P}_{\Omega_\infty}
 = \int_{\Omega_\infty} \rho ({\bf r},\infty) d{\bf r}
 = \lim_{t \to \infty} \int_{\Omega_t}
       \rho ({\bf r},t) d{\bf r}
 = \lim_{t \to \infty} \mathcal{P}_{\Omega_t}(t) .
 \label{ec7}
\ee
The difference of this expression with respect to Eq.~(\ref{ec6})
is that now we can keep track of the amount of probability going
into $\mathcal{D}=\Omega_\infty$ by means of an unambiguous
time-dependent relationship. But, since the probability inside the
corresponding tube remains constant, we can just write down
(\ref{ec7}) as
\be
 \mathcal{P}_{\Omega_\infty}
   = \int_{\Omega_0} \rho({\bf r},0) d{\bf r}
   = \mathcal{P}_{\Omega_0}(0) .
 \label{ec8}
\ee
The initial restricted probability $\mathcal{P}_{\Omega_0}$ can be
computed from an appropriate sampling of (Bohmian) initial
conditions (according to $\rho({\bf r},0)$) inside $\Omega_0$, as in
classical mechanics. By proceeding in this way, the physical meaning
of Eq.~(\ref{ec8}) becomes more apparent (and almost trivial): given
a certain set of initial conditions enclosed in some region of the
configuration space, their total number is conserved regardless of
how the ensemble evolves. This is a result of {\it general}
validity, which goes again beyond standard quantum mechanics, for it
states that the probability within a certain region of the
configuration space can be transported to another one causally
connected, i.e., in an unambiguous fashion when following
probability tubes.


\section{Numerical simulations}
\label{sec3}


\subsection{Tunneling}
\label{sec31}

To test the feasibility of the concepts exposed in Sec.~\ref{sec2}
and, in particular, the applicability of Eq.~(\ref{ec8}) regardless
of the initial state or the problem considered, consider the
scattering of a wave function off a barrier. This tunneling problem
may describe, for example, the passage from reactants to products in
a chemical reaction. Thus, let us consider as the initial wave
function an arbitrary coherent superposition of three Gaussian wave
packets,
\be
 \Psi_0(x) = A_0 \sum_{i=1}^3 c_i \psi_i(x) ,
 \label{init1}
\ee
where each $\psi_i$ is described by
\be
 \psi_0(x) = \left(\frac{1}{2\pi\sigma_0^2}\right)^{-1/4}
  e^{- (x-x_0)^2/4\sigma_0^2 + i p_0 (x - x_0)/\hbar} ,
 \label{eq1}
\ee
with $x_0$ and $p_0$ being respectively the (initial) position and
momentum of the wave-packet centroid (i.e., $\langle \hat{x} \rangle
= x_0$ and $\langle \hat{p} \rangle = p_0$), and $\sigma_0$ its
initial spreading. Without loss of generality, the parameters chosen
are: $(c_1,c_2,c_3) = (1.0,0.75,0.5)$, $(x_{0,1},x_{0,2},x_{0,3}) =
(-10,-12,-9)$, $(p_{0,1},p_{0,2},p_{0,3}) = (10,20,15)$,
$(\sigma_{0,1},\sigma_{0,2},\sigma_{0,3}) = (0.2,1.6,0.8)$, and $m =
\hbar = 1$; after introducing these values, the wave function
$\Psi_0$ is properly renormalized before starting the simulation,
this being denoted by the constant prefactor $A_0$ in (\ref{init1}).
As for the barrier, we take a nearly square barrier consisting of a
sum of two hyperbolic tangents,
\be
 V(x) = \frac{V_0}{2} \left\{ \tanh \left[ \alpha (x - x_-) \right]
  - \tanh \left[ \alpha (x - x_+) \right] \right\} ,
 \label{pot}
\ee
with $V_0 = 150$, $\alpha = 10$, and $x_\pm = \pm 2$. The simulation
was carried out by means of a standard wave-packet propagation
method
\cite{leforestier:jcompphys:1991,kosloff:JCompPhys:1983,kosloff:JCP:1983},
while the associated Bohmian trajectories were computed
``on-the-fly'', substituting the wave function resulting at each
iteration into (\ref{ec6}) and then integrating this equation of
motion.

\begin{figure}[t]
\begin{center}
 \includegraphics[width=10cm]{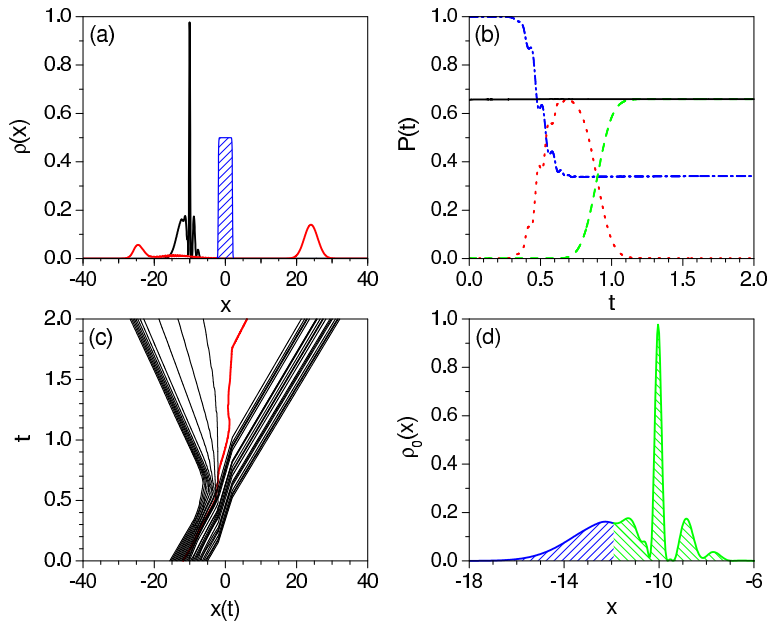}
 \caption{\label{fig1}
  (a) Initial (black line) and final (red line) probability densities
  in the problem of scattering off a nearly square barrier (blue
  shadowed region).
  (b) Time-dependence of the transmission (green dashed line),
  reflection (blue dash-dotted line) and intra-barrier (red dotted
  line) probabilities.
  The probability enclosed in $\Omega_t$ and obtained with the aid of the
  Bohmian calculation is displayed with black solid line.
  (c) Bohmian trajectories illustrating the process dynamics;
  the separatrix is denoted with red thicker line.
  (d) Splitting of the initial probability density according to the
  separatrix initial position.
  Only the green shadowed region ($\Omega_0$) contributes to
  transmission.}
\end{center}
\end{figure}

The initial and final probability densities are displayed in
Fig.~\ref{fig1}(a) (black and red solid lines, respectively); the
barrier has also been plotted (blue shadowed region). The system
wave function was evolved until the probability within the
intra-barrier region was negligible, this being assumed to be our
asymptotic time. This can be better seen in panel (b), where the
transmission (green dashed line), reflection (blue dash-dotted line)
and intra-barrier (red dotted line) probabilities are monitored
along time. In the calculation of these restricted probabilities it
was assumed that: $\mathcal{D}_{\rm T}$ is the region beyond the
right-most barrier edge, $\mathcal{D}_{\rm I}$ is the region
confined between the two barrier edges, and $\mathcal{D}_{\rm R}$ is
the region to the left-most edge. As seen in the figure, after $t
\approx 1.15$, we find that $\mathcal{P}_{\rm R} \approx 0$ and
$\mathcal{P}_{\rm T}$ reaches its maximum (asymptotic) value, which
already remains constant with time.

Let us split up the initial wave function into ``reflectable'' and
``transmittable'', with $\Omega_0$ encompassing the portion
associated with the latter. The upper bound for this region can be
the initial position, $x(0)$, of any trajectory on the right-most
border of the initial probability density, for which $\rho(x(0))
\approx 0$. For the lower boundary, a search has to be done
\cite{sanz:JPA:2011}, so that it is ensured that the chosen
trajectory is the last (or nearly the last) one in crossing the
right-most barrier edge and not displaying a backwards motion.
Determining this trajectory constitutes a major drawback, since
Bohmian trajectories are not analytical in general, neither there is
a simple, general way to make an estimate \cite{sanz:JPA:2011}. This
initial condition has to be then determined either from a series of
sampling runs or just fixing the asymptotic value of the trajectory
and running backwards in time the dynamics until $t=0$.

A sampling set of Bohmian trajectories is shown in
Fig.~\ref{fig1}(c), with their initial positions evenly distributed
along the extension covered by the initial probability density. The
red thicker line denotes the separatrix splitting the initial swarm
into two groups of trajectories: those that will surmount the
barrier (transmitted) and those that will bounce backwards
(reflected). Accordingly, at any time $t$, the region $\Omega_t$
(i.e., the time-evolved of $\Omega_0$) always confines trajectories
that eventually become transmitted; $\partial\Sigma_t$ is determined
by the positions at $t$ of two trajectories, namely the separatrix
and the rightmost one considered. The evolution of these two
trajectories defines the corresponding transmission probability
tube, along which all the transmitted probability density flows.

Bearing this scheme in mind, it is now rather simple and
straightforward to determine which part of the initial probability
density contributes to tunneling transmission, denoted by the green
shadowed area in panel (d). The integral over this area readily
provides the value otherwise found from the asymptotic
$\mathcal{P}_{\rm T}$ (see panel (b)). Actually, the evaluation of
$\mathcal{P}_{\Omega_t} (t)$ at each time renders a constant value
(see black solid line in panel (b)), thus proving the conservation
of the probability inside $\Omega_t$. Furthermore, this also proves
our assertion that final probabilities can be, in principle,
directly and unambiguously obtained from the initial state by means
of Bohmian trajectories.


\subsection{Grating diffraction}
\label{sec32}

Consider now that we would like to determine, for example, the
so-called {\it peak-intensity area} in a scattering problem, i.e.,
the total intensity that goes into a certain diffraction peak or,
equivalently, the relative amount of scattered particles lying
within the area covered by this diffraction peak. This value is
obtained by computing the integral of the probability density lying
between the two adjacent minima associated with such a diffraction
peak. One could be legitimated to ask about which parts of the
diffracted beam contribute to a particular diffraction peak, or how
each particular feature of the diffracting object contributes to a
final diffraction feature. These questions cannot be answered within
the standard version of quantum mechanics. However, the tools so far
developed can be used to provide reasonable answers (i.e., coherent
with the use of a full quantum framework), in particular
Eq.~(\ref{ec8}).

\begin{figure}[t]
\begin{center}
 \includegraphics[width=15cm]{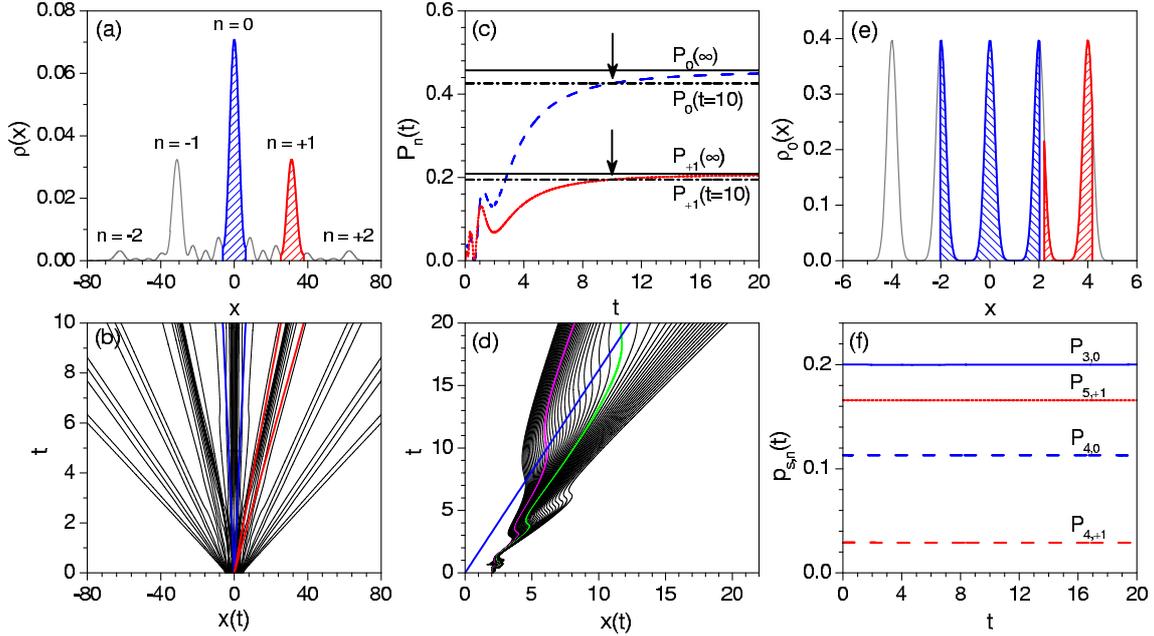}
 \caption{\label{fig2}
  (a) Five-slit diffraction pattern at $t=10$.
  Principal maxima are labeled according to their diffraction order,
  $n$; shaded color areas denote the computed peak-intensity areas.
  (b) Bohmian trajectories illustrating the diffraction dynamics.
  The domains $\mathcal{D}_0$ and $\mathcal{D}_{+1}$ for the maxima
  highlighted in (a) are enclosed by the boundaries $x_0^\pm$ (blue)
  and $x_{+1}^\pm$ (red), respectively (see text for details).
  (c) Peak-intensity areas corresponding to the diffraction orders
  $n=0$ (blue dashed line) and $n=+1$ (red dotted line).
  The peak-intensity areas directly computed from the initial state
  are denoted with black solid line for $t=20$ (asymptotic time) and
  black dash-dotted line for $t=10$.
  (d) Set of Bohmian trajectories around the separatrix for the peak
  $n=0$ at $t=10$ (purple thicker line).
  The separatrix at $t=20$ and the boundary $x_0^+$ (see text for
  details) are represented with green and blue thicker lines,
  respectively.
  (e) Sections of the initial probability density that give rise to
  the $n=0$ (blue shaded section) and $n=+1$ (red shaded section)
  diffraction peaks at $t=10$ (see panel (a)).
  (f) Contributions from slits 3, 4 and 5 to the $n=0$ and $n=+1$
  diffraction peaks (see panel (d)).}
\end{center}
\end{figure}

To illustrate this situation, we are going to analyze five-slit
diffraction \cite{sanz:JPCM:2002}, which is also a good example to
observe branching processes and bifurcations of the probability
tubes. We shall consider our initial wave function to be the
diffracted beam. Let us then assume that spatial variations along
the direction perpendicular to the grating are negligible compared
to the transversal ones and the propagation (along that direction)
is faster. This working hypothesis allows us to characterize the
process by a wave function that only accounts for what happens in
the transversal direction. If we also assume slit Gaussian
transmission, the initial wave function can be expressed as a
coherent superposition of five Gaussian wave packets,
\be
 \Psi_0(x) = A_0 \sum_{i=1}^5 \psi_(x) ,
\ee
where each wave packet is given by (\ref{eq1}), with $x_0^{(i)} = -
4 + 2(i-1)$, $p_0^{(i)} = p_0 = 0$ (zero transverse momentum),
$\sigma_0^{(i)} = \sigma_0 = 0.2$, and $m = \hbar = 1$; as before,
$A_0$ is the renormalization constant. Regarding the numerical
simulation, we have followed the same procedure as in
Sec.~\ref{sec31}. Propagation up to $t=10$ produces diffraction
peaks that are already well resolved, but with nonzero adjacent
minima, as seen in Fig.~\ref{fig2}(a). This can be the case, for
example, when the detector is allocated relatively close to the
grating, so that the system has not been able yet to reach the
Fraunhofer regime \cite{sanz:prb:2000}. The corresponding Bohmian
dynamics is illustrated with the trajectories displayed in panel
(b), with initial conditions evenly distributed along the effective
extension covered by each wave packet.

Consider the principal maxima associated with the diffraction orders
$n=0$ and $n=+1$ in panel (a). Let $\Omega_0 \equiv
\{x_{0}^{-},x_{0}^+\}$ and $\Omega_{+1} \equiv
\{x_{+1}^{-},x_{+1}^+\}$ be the sections of the initial wave
function, such that at $t=10$ cover the principal maxima $n=0$ and
$n=+1$, respectively, between their corresponding adjacent minima
(see color shaded regions in the figure). The restricted
probabilities associated with these two regions give us the value of
the respective peak-intensity areas. From a standard quantum
viewpoint, one could be tempted to define some associated sectors or
domains, $\mathcal{D}_n$, taking into account the Fraunhofer limit,
where minima vanish. More specifically, the domain $\mathcal{D}_n$
for the $n$th diffraction peak would cover the extension between the
corresponding adjacent minima. According to the Fraunhofer
diffraction formula for this case \cite{sanz:JCP-Talbot:2007}, the
boundaries for $\mathcal{D}_n$ evolve with time\footnote{It does not
make any sense to consider in this case of fixed domain, because the
wave function is spreading all the way through ---even in the
Fraunhofer regime, where this spreading is linear with time,
although the relative shape remains invariant.} as
\be
 x_n^\pm(t) = 2\pi(N \pm 1) \left( \frac{n}{N} \right)
  \left( \frac{\hbar}{md} \right) t ,
 \label{clbound}
\ee
with $N=5$ and $d=2$, and where $+$/$-$ refers to the right/left
boundary. Notice that this expression is valid within the Fraunhofer
regime, where minima vanish; for shorter times, these boundaries
will pass through positions with nonzero probability density and, at
$t=0$, all boundaries will coalesce on $x=0$. In panel (b) the
boundaries for the diffraction orders $n=0$ and $n=+1$ are shown
(straight trajectories of the same color as the corresponding shaded
areas from panel (a)).

From a practical viewpoint, computing restricted probabilities with
the aid of these domains $\mathcal{D}_n$ thus requires to be in the
Fraunhofer regime, which is troublesome in the sense that this
condition is not always fulfilled. This is precisely what we observe
in panel (c), where the restricted probabilities within the domains
$\mathcal{D}_0$ (blue dashed line) and $\mathcal{D}_{+1}$ (red
dotted line) have been computed. Asymptotically, they approach a
constant value, which has been obtained by means of proper Bohmian
boundaries (see below). Thus, at $t=20$ the deviations with respect
to these asymptotic values are relatively small (about 1.72\% for
$n=0$ and 1.68\% for $n=+1$). However, for smaller values of time,
discrepancies become more relevant. The problem with these domains
is that there is no way to determine the origin of these divergences
(other than the lack of validity of expression (\ref{clbound})) and
therefore to control them.

To understand such a behavior, we need to consider Bohmian
trajectories and have a look at their topology. In panel (d) we show
a swarm of trajectories with very close and evenly distributed
initial conditions (they cover 0.030 space units). As seen, these
trajectories start on the right side of $x_0^+$; as time proceeds,
some of them start crossing this boundary; eventually, they split
into two groups, each one contributing to a different diffraction
peak, namely $n=0$ and $n=+1$, in spite of their initial proximity.
This is a clear example of branching processes, as mentioned in
Sec.~\ref{sec2}. In these cases where we are not yet in the
Fraunhofer domain, it is thus important to make clear what the
asymptotic time is, since the separatrix at one time can be useless
at another time. For example, at $t=0$ the separatrix is denoted by
the purple thicker line in panel (d), which renders restricted
probabilities for $n=0$ and $n=+1$ about 6.92\% and 6.70\% lower
than the corresponding asymptotic ones (see black solid lines in
panel (c)). This means that about 6.92\% and 6.70\% of trajectories
are still lacking in the calculation of the corresponding peak
areas, as it is inferred by looking at the separatrix denoted with
the green thicker line in panel (d).

Finally, given the presence of branching, one may be interested in
determining how much each slit contributes to the final pattern,
also from the initial state. This can be easily done with the aid of
the separatrix trajectories, which allow us to establish which is
the range of the initial wave function contributing to each
diffraction peak and therefore the source slit. Thus, in panel (e)
we observe that the whole central slit and more than a half of the
adjacent ones contribute to the $n=0$ diffraction peak (see blue
shadowed areas), while a small portion of the third slit and a large
one of the fifth slit contribute to $n=+1$ (see red shadowed areas).
By integrating these portions of initial wave function, we can now
provide a quantitative measure of the relative contribution of each
slit, as shown in panel (f). In this figure, $\mathcal{P}_{i,n}$
refers to the restricted probability contributing to the $n$th
diffraction peak and coming from the $i$th slit.


\section{Concluding remarks}
\label{sec4}

We have shown how final features of quantum probabilities can be
unambiguously related to different sections of the initial state by
using Bohmian mechanics. This fact certainly reminds of some
classical statistical approaches that can be found in different
fields, e.g., chemical reactivity
\cite{pollak:JCP:1980,pollak:JCP:1983,tannor-bk} or atmospheric
modeling \cite{egger:MonWeaRev:1996,sommer:MonWeaRev:2010}. More
specifically, such a connection is enabled through the combination
of the divergence or Gauss-Ostrogradsky theorem with Bohmian
mechanics, which allows to go a step beyond the standard formulation
of quantum mechanics and to devise new tools to study and to understand
the physics underlying of quantum systems. In particular, even if
not observable from a experimental point of view, the fact that one
can properly define (in terms of separatrices) regions that are
uniquely transported throughout the system configuration space
without loss or gain of quantum flux results of much interest to
study the dynamical role of quantum phase in time-resolved
experiments. Note that this methodological prescription allows to
monitor the detailed evolution of a set of particular initial
conditions and, therefore, to determine the final outcome
(probability) directly from the initial state. In this sense, it can
be considered as an extension to time-dependence of Born's rule
(see Appendix below).
Actually, even if one does not know where such a particular set or
part of the initial wave function will evolve, it is sure that the
probability confined within its boundaries (separatrices) will not
mix up with contributions from other parts of the same initial wave
function. Within the standard scenario, however, there is no
certainty about this and hence one has to appeal to rather arbitrary
(classical or semiclassical) methods and/or arguments to determine
final restricted probabilities.

A criticism that could be risen against this approach, though, is
that the Bohmian probability tubes depend on the initial wave
function (they are {\it context dependent}) and therefore varying
the latter unavoidably leads to a change of the topology of the
tubes (even for the same physical problem). For example, consider
the case of tunneling analyzed above. Once the transmitted part is
known, one can determine the region of the initial state that gives
rise to such a transmission. However, the same tube cannot be used
to analyze a new initial wave function.
More specifically, in the context of concrete computations, the
choice of the initial wave function will conform with some physical
limitations, e.g., it is meaningless to consider an initial wave
function not localized around the domain of initial conditions whose
transmission is studied.
Now, since the topology displayed by Bohmian trajectories is sensitive
to the particular form adopted by the initial wave function, it is
clear that such limitations are also going to restrict the relevance
of contextual issues from a practical (computational) viewpoint.
That is, by limiting the number of possible initial states, feasible
boundaries for the corresponding probability tubes are also being
somehow determined.
Therefore, even if the true boundary is not known with a high accuracy,
at least one has a fair estimate of it, which can be used for practical
purposes \cite{sanz:JPA:2011}.
Nonetheless, in any case, it is important to stress that this contextual
dependence is not a problem of the approach itself (nor of Bohmian
mechanics, generally speaking), but a property of quantum mechanics,
which manifests more remarkably through Bohmian mechanics. Any
quantum outcome is thus strongly dependent on the initial ensemble
considered (i.e., the whole wave function), contrary to what we find
in classical mechanics when considering trajectories in phase space,
where the particle distribution does not influence the final outcome
(and therefore the definition of classical tubes). This
context-dependence appears, for example, in quantum control schemes
\cite{brumer-bk:2003}, which are based on this property: by
manipulating the initial state in a particular way, we can inhibit
or enhance a certain final property.

There is another important issue worth stressing, which could also
be considered as a drawback. It arises when dealing with chaotic
and/or large systems, and is common to any trajectory-based
methodology, including classical ones. In such cases, one only knows
where the trajectories go after completion of the simulation. Thus
information about the system dynamics is always extracted a
posteriori, which is useful, though, to interpret reaction
probabilities or momentum/energy transfers solely. In such cases,
notice that multi-dimensional tubes with multiple branches may
appear, thus making relatively complex the analysis of the system
under study. Even though, such an analysis is still possible and
useful, because it allows us to classify sets of initial conditions
according to the type of dynamics that they will lead to.

The approach posed here, in principle, could be used to compute
probabilities without any need for solving the full dynamics of the
process, but only with some knowledge provided by the quantum
trajectories. In practice, due to the non-analyticity of Bohmian
trajectories, this cannot be easily done. However, it should be
mention that in the literature somehow related methods can be found,
which operate the other way around, i.e., from probability densities
they try to infer the corresponding quantum trajectories without
solving the associated equations of motion. This is the case, for
example, of the Bohmian Monte Carlo sampling method
\cite{coffey:JPA:2008}, based on the idea of quantile motion
\cite{brandt:PLA:1998}, or the kinematic approach
\cite{coffey:JPA:2010}, based on Voronoi's tessellations method
\cite{fonseca-guerra:JComputChem:2004}.


\section*{Acknowledgements}

Support from the Ministerio de Econom{\'\i}a y Competitividad
(Spain) under Projects and FIS2010-22082 and FIS2011-29596-C02-01,
as well as from the COST Action MP1006 ({\it Fundamental Problems in
Quantum Physics}) is acknowledged. The authors are grateful to
anonymous referees for useful comments. A. S. Sanz thanks the
Ministerio de Econom{\'\i}a y Competitividad for a ``Ram\'on y
Cajal'' Research Fellowship and the University College London for
its kind hospitality.


\appendix
\section*{Appendix. Bohm-Born rule}

\renewcommand{\theequation}{A.\arabic{equation}}
\setcounter{equation}{0}

The results discussed in Sec.~\ref{sec2} lead us straightforwardly
to establish a connection with the so-called Born rule
\cite{born:ZPhys:1926,zurek-bk,landsman:CQP:2009,brumer:pra:2006}.
Actually, the aforementioned combination of the quantum continuity
equation and Bohmian mechanics makes Eq.~(\ref{ec8}) to implicitly
contain a kind of time-dependent Born rule. This is readily seen as
follows. Consider two arbitrary times, $t$ and $t'$ (we will assume
$t' > t$). It follows from Eq.~(\ref{ec8}) that
\be
 \int_{{\bf r}(t')} \rho[{\bf r}(t')] d{\bf r}(t') =
  \int_{{\bf r}(t)} \rho[{\bf r}(t)] d{\bf r}(t) .
 \label{ec8bis}
\ee
On the other hand, because of the causal connection between ${\bf
r}(t)$ and ${\bf r}(t')$ in Bohmian mechanics, one can also define a
Jacobian
\be
 \mathcal{J}[{\bf r}(t)] =
  \frac{\partial {\bf r}(t')}{\partial {\bf r}(t)} ,
 \label{eq23bbis}
\ee
which describes the mapping transformation in configuration space
from $x(t)$ at a time $t$ to $x(t')$ at a time $t'$. This relation
is equivalent to the one found in classical mechanics when solving
the (classical) continuity equation
\cite{gutzwiller:1990,sanz:SSR:2004,sanz:PhysRep:2007}, although in
that case it includes the corresponding momenta, since it is defined
in phase space. Thus, taking into account the equality
(\ref{ec8bis}) and the connection between the layers defined by
$d{\bf r}(t)$ and $d{\bf r}(t')$ enabled by the Jacobian,
\be
 d{\bf r}(t') = | \mathcal{J}[{\bf r}(t)] |\ \! d{\bf r}(t) ,
 \label{eq23babis}
\ee
the probability density evaluated along a quantum trajectory at a
time $t'$ is related through the inverse Jacobian transformation
with its value at an earlier time $t$, as
\be
 \rho[{\bf r}(t')] = | \mathcal{J}[{\bf r}(t')] |\ \! \rho[{\bf r}(t)] ,
 \label{eq23bbbis}
\ee
with $| \mathcal{J}[{\bf r}(t')] | = | \mathcal{J}[{\bf r}(t)]
|^{-1}$. That is, Born's rule is preserved along time whenever the
evolution of the probability $\rho[{\bf r}(t)]$ is monitored within
the probability tube defined by the swarms of quantum trajectories
${\bf r}(t)$ and ${\bf r}'(t) = [{\bf r} + d{\bf r}](t)$:
\be
 \rho[{\bf r}(t')] d{\bf r}(t') = \rho[{\bf r}(t)] d{\bf r}(t) .
 \label{ec8bbis}
\ee




\end{document}